\documentstyle[preprint,prc,aps]{revtex}
\begin{document}
\draft

\title{Combined analysis of the reactions $pp \to pp$, $\pi d\to \pi d$, and 
$\pi d\to pp$} 

\author{Chang-Heon Oh, Richard A. Arndt, Igor I. Strakovsky\cite{iis}, 
and Ron L. Workman}
\address{Department of Physics, Virginia Polytechnic Institute and State
University, Blacksburg, VA 24061}

\date{\today}

\maketitle

\begin{abstract}
Results are presented for a combined analysis of the reactions $pp\to pp$,
$\pi d\to \pi d$ and $\pi d\to pp$ over the $\sqrt{s}$ interval from
pion threshold to approximately 2.4~GeV. 
These results for $\pi d\to pp$ and $\pi d$ elastic
scattering are superior to our previous analyses of these reactions. In
particular, the overall phase in $\pi d\to pp$ has now been determined. 
Comparisons are made with previous (separate and combined) analyses of this
two-nucleon system. 
\end{abstract}

\vspace*{0.5in}


\vspace*{0.5in}

\narrowtext

\section{INTRODUCTION}

An understanding of the $NN$ interaction is fundamental
to studies of the more general $\pi NN$ problem\cite{Mizutani}.  
Below 1~GeV, in proton laboratory kinetic energy $T_{p}$ for the $NN$ system, 
the dominant channels contributing to $NN$ inelasticity are 
$\pi d$ and $N\Delta$\cite{NDelta}.  At these energies, it 
is useful to employ a multi-channel formalism in analyzing 
all existing data simultaneously.
In the present work, we have used the K-matrix formalism in order to unify 
the analysis of several reactions ($pp\to pp$\cite{pp}, 
$\pi d\to \pi d$\cite{pid}, and $\pi d\to pp$\cite{pidpp}) which we 
have, in the past, considered separately. The range of $\sqrt{s}$ was 
chosen to include all of our results for the pion-induced reactions 
($T_{\pi}=0-500~MeV$).   

Clearly, we are not the first to consider this problem. 
A joint analysis of these three reactions, in a narrow energy range 
near the $N\Delta$ threshold, was recently reported by
Nagata et al.\cite{Nagata}. This work used a mix of model-based and
phenomenological results to investigate possible narrow structures in
these reactions. An older work by Edwards\cite{Edwards} used the
multi-channel K-matrix formalism to study the $J^P=2^+$ and $3^-$ 
states associated with dibaryon candidates. 

The present analysis differs from those carried out previously in a number of
important respects. We did not restrict our study to partial-waves
containing interesting structures. For $pp$ elastic scattering, all waves
with $J\le 7$ were used. Partial wave with $J\le 5$ were retained for both
$\pi d$ elastic scattering and $\pi d\to pp$.
In addition, the K-matrix parameters were determined solely from our
fits to the available data bases for each separate reaction. 
No results of outside analyses or any model approaches were used 
as constraints.  As a result, the amplitudes found in our K-matrix fits are
as ``unbiased" as those coming from the separate analyses\cite{said}.

In Section~II, we will outline the K-matrix formalism used in this analysis.
The combined and separate analyses will
be compared in Section~III. Conclusions and suggestions for further study
will be given in Section~IV.

\section{FORMALISM}     

In order to analyze the reaction $\pi d\to pp$ along with elastic 
$pp$ and $\pi d$ scattering, we have constructed a K-matrix formalism
having $pp$, $\pi d$ and $N\Delta$ channels. 
The energy-dependence of our global fit was obtained through a coupled-channel 
K-matrix form in order to ensure that unitarity would not be violated. 
The ``$N\Delta$" channel is
added to account for all channels other than $pp$ and $\pi d$. The 
most important thresholds are illustrated schematically in Fig.~1.
That this catch-all channel is indeed mainly $N\Delta$ can be seen in
Fig.~2, where the total cross sections for $pp$ and $\pi d$ scattering 
are broken into their components.  

As the elastic $pp$ partial-wave analysis is far superior to the 
$\pi d$ elastic and $\pi d\to pp$ analyses, we have carried out fits
in which the $pp$ partial-waves were held fixed.  (The partial wave 
decomposition the of $pp$, $\pi d$, and $N \Delta$ systems are given 
in Table~I.) As described below, the 
$pp$ amplitudes were used to fix some elements of the K-matrix, while
the others were determined from a fit to the combined $\pi d$ elastic
and $\pi d\to pp$ data bases. 

States of a given total angular momentum and parity ($J^P$) were 
parameterized by a 4x4 K-matrix ($K_J$) which coupled
to an appropriate $N\Delta$ channel. Spin-mixed(2x2) $pp$ states couple
to unmixed $\pi d$ states, and unmixed $pp$ states couple to 
spin-mixed(2x2) $\pi d$ states, so the $\pi d - pp$ system is always 
represented by a 3x3 matrix. For example, the T-matrix ($T_J$) for 
$J^P=2^+$ (unmixed $pp$ states) is given by \\
\hspace*{2.28in} $pp$ \hspace*{.23in} $\pi d_-$ \hspace*{.15in} $\pi d_+$
\begin{equation}
{T_2 =
\left( \begin{array}{lll}
^1D_2  & ^1D_2P     & ^1D_2F     \\
^1D_2P & ^3P_2      & \epsilon_2 \\
^1D_2F & \epsilon_2 & ^3F_2
\end{array} \right)}
{\begin{array}{l}
\qquad pp \\ \qquad \pi d_- \\ \qquad \pi d_+
\end{array}} 
\end{equation}
\newpage
whereas the T-matrix for $J^P=2^-$ (mixed $pp$ states) is \\
\hspace*{2.28in} $pp_-$ \hspace*{.15in} $pp_+$ \hspace*{.15in} $\pi d$
\begin{equation}
{T_2 =
\left( \begin{array}{lll}
^3P_2      & \epsilon_2 & ^3P_2D \\
\epsilon_2 & ^3F_2      & ^3F_2D \\
^3P_2D     & ^3F_2D     & ^3D_2
\end{array} \right)}
{\begin{array}{l}
\qquad pp_- \\ \qquad pp_+ \\ \qquad \pi d
\end{array}} 
\end{equation}
The subscripts $\pm$ denote states with $L=J\pm1$.
In the above, the mixing parameters ($\epsilon$) for elastic $pp$ and
$\pi d$ scattering are different. For the reaction $\pi d\to pp$, the
notation ($^{2S_{pp} + 1}L^{pp}_J L^{\pi}$) of Ref.\cite{pidpp} is used.

Adding an $N\Delta$ channel results in a 4x4 T-matrix. Dropping the 
$J$-subscript, we write the K-matrix as
\begin{equation}
K = \left( \begin{array}{ll}
  K_{pp}     &  K_0  \\
  \tilde K_0 & K_i
  \end{array} \right) ,
\end{equation}
where $K_{pp}$ is the elastic $pp$ scattering sub-matrix, $K_0$ and 
$\tilde K_0$ are row and column vectors, and $K_i$ is the sub-matrix of
channels involving $\pi d$ and $N\Delta$ states. This K-matrix can be
re-expressed as a T-matrix
\begin{equation}
T = \left( \begin{array}{ll}
  T_{pp}     &  T_0  \\
  \tilde T_0 & T_i
  \end{array} \right) 
\end{equation}
using the relation T=K(1--iK)$^{-1}$. We then have the correspondence
\begin{equation}
T_{pp} = \bar K_{pp} (1 - i \bar K_{pp} )^{-1} ,
\end{equation}
where
\begin{equation}
\bar K_{pp} = K_{pp} + i K_0 (1 - i K_i)^{-1} \tilde K_0
\end{equation}
In order to ensure an exact fit to the $pp$ elastic T-matrix, given by
our most recent analysis of $NN$ elastic scattering to 
1.6~GeV\cite{pp}, we take
\begin{equation}
K_{pp} = T_{pp} (1 + i T_{pp} ) - i K_0 (1 - i K_i )^{-1} \tilde K_0.
\end{equation}
The matrix elements are then expanded as polynomials
in the pion energy times appropriate phase-space factors. The $\pi d$
elastic and $\pi d\to pp$ T-matrix elements are extracted from
$T_0$ and $T_i$.

\section{PARTIAL-WAVE AMPLITUDES}

We have fitted the amplitudes for $pp\to pp$ and the existing data bases 
for $\pi d\to pp$, and $\pi d\to \pi d$, 
using the K-matrix formalism outlined in Section~II.  
The $\pi d$ elastic and $\pi d\to \pi d$ data bases
used in this analysis are described in Refs.\cite{pid} and \cite{pidpp},
and available from the authors\cite{said}. 
The overall $\chi^2$ for our combined analysis is actually superior to that 
found in our single-reaction analyses. This is due to the improved 
parameterization scheme. A comparison is given in Table~II.
We should emphasize that the amplitudes for $pp$ 
elastic scattering are the same as those given in Ref.\cite{pp}.  
As mentioned above, this feature was built into our K-matrix 
parameterization. For this reason, we have omitted plots of the 
$pp$ amplitudes\cite{said}.  

The results for $\pi d$ elastic scattering
are also qualitatively similar, up to the limit of our single-energy
analyses. In Fig.~3 we compare the main partial-waves from our 
single-reaction analysis\cite{pid} and combined analysis (solution C500). 
Significant differences begin to appear above a pion laboratory kinetic 
energy of 300~MeV or 2.3~GeV in $\sqrt{s}$.  
(The $^3D_2$ partial wave from C500 is an exception, 
departing from the single-reaction analysis near threshold.) The upper
limit to our single-energy analyses is due to a sharp cutoff in the number
of data. This is apparent in Fig.~2 of Ref.\cite{pid}. 
Much additional data above 300~MeV will be required before a 
stable solution to 500~MeV can be expected. 

A comparison of results for 
$\pi d\to pp$ reveals the most pronounced differences. 
One reason for this is the overall phase
which was left undetermined in Ref.\cite{pidpp}. There, we arbitrarily 
chose the $^3P_1S$ wave to be real. In the present analysis, the overall
phase has been determined. In Fig.~4 we show that the $^3P_1S$ phase is 
very different in the combined and separate analyses. 
Given the large difference in overall phase, we have chosen to compare the
partial-wave amplitudes from the separate and combined analyses in terms of 
their moduli.  This comparison is made in Fig.~5.  As was the case for
$\pi d$ elastic scattering, differences are most significant above
approximately 2.3~GeV in $\sqrt{s}$. A similar lack of data exists above
this energy.  

In general we see a good agreement for the dominant amplitudes
found in the separate and combined analyses. 
Figures 3 and 5 also display our single-energy analyses which were  
done in order to search for structure which may be 
missing from the energy-dependent fit.  (Details of the single-energy 
analyses are given in Refs.\cite{pid,pidpp}.) A comparison of the 
single-energy and energy-dependent fits is given in Tables~III and IV.

\section{SUMMARY AND CONCLUSIONS}

We have obtained new partial-wave amplitudes for $\pi d$ elastic 
scattering and the reaction $\pi d\to pp$, using a K-matrix method
which utilized information from our elastic $pp$ scattering analysis.
In addition to producing amplitudes more tightly constrained by 
unitarity, we have resolved the overall phase ambiguity existing
in our previous $\pi d\to pp$ analysis.

As mentioned in Section III, the combined analysis has resulted in a 
slightly improved fit to the $\pi d$ elastic and $\pi d \to pp$ 
data bases. The most noticeable differences, at the partial-wave level,
appear at higher energies where the existing data are sparse. It
is difficult to find cases where the fit has been dramatically improved.
One exception is the set of $\pi d$ total cross section data between
300 and 500~MeV. Here the combined analysis is much more successful in
reproducing the energy dependence. The combined analysis gives total
cross sections which begin to rise at 500~MeV, whereas the
separate analysis shows a fairly monotonic decrease from 400 to 500~MeV.
The behavior seen in the combined analysis seems reasonable, as
the $\pi d$ total cross sections do begin to rise just beyond the upper
energy limit of our analysis. Many of the individual partial-wave 
amplitudes from C500 show rising imaginary parts near 500~MeV, 
a feature absent in the analysis of $\pi d$ elastic data alone. 

The present analysis has also resulted in a unified description of the
resonancelike behavior previously noted in our separate analyses of 
$pp$\cite{pp} and $\pi d$\cite{pid} elastic scattering, 
and the reaction $\pi d\to pp$\cite{pidpp}. This behavior\cite{review} has 
been variously described as ``resonant'' (due to the creation of dibaryon
resonances) and ``pseudo-resonant'' (due to the $N\Delta$ intermediate 
state).  We expect that our combined analysis will further constrain models
based on these two mechanisms.

\vskip .2cm
 
\acknowledgments

I.~S. acknowledges the hospitality extended by the Physics Department of
Virginia Tech.
This work was supported in part by a U.~S. Department of
Energy Grant DE-FG05-88ER40454.

\eject


\eject

{\Large\bf Figure captions}\\
\newcounter{fig}
\begin{list}{Figure \arabic{fig}.}
{\usecounter{fig}\setlength{\rightmargin}{\leftmargin}}
\item
{Energy scale in terms of the total center-of-mass energy  
($\sqrt{s}$) and the incident kinetic energies of the $pp$ (T$_{p}$) and 
$\pi d$ (T$_{\pi}$) initial states. 
The locations of relevant thresholds are also displayed.}
\item
{Total cross sections $\sigma _{tot}$ (solid) and total elastic cross sections 
$\sigma_{el}$ (dashed) correspond to the C500 solution.  Data for 
$\sigma _{tot}$ (open circles) are taken from the SAID database \cite{said}.  
(a)
Dash-dotted lines, corresponding to the C500 solution, show the total 
cross sections ($\sigma _{\pi d}$) for $pp \to \pi d$. 
The corresponding data 
from the SAID database \cite{said} are plotted as open triangles. 
The remainder ($\Delta \sigma$) 
is given by $\sigma _{tot} - \sigma _{el} - \sigma _{\pi d}$ and plotted
as a dotted line. 
Total cross sections for the reactions 
$pp \to \Delta ^{+} p + \Delta ^{++} n$ \cite{NDelta} are plotted as dark 
circles. 
(b)
Dash-dotted lines (C500) show the total 
cross sections ($\sigma _{pp}$) for $\pi d\to pp$. 
The corresponding data 
from the SAID database \cite{said} are plotted as open triangles. 
The remainder ($\Delta \sigma$) is
given by $\sigma _{tot} - \sigma _{el} - \sigma_{pp}$ and plotted 
as a dotted line.}  
\item
{Partial-wave amplitudes of the reaction 
$\pi d\to \pi d$ from T$_{\pi}$ = 0 to  500~MeV.  
Solid (dashed) curves give the real (imaginary) parts of amplitudes 
corresponding to the C500 solution.  Our previous analysis (SM94)  
\protect\cite{pid} is plotted with long dash-dotted (real part) and short 
dash-dotted (imaginary part) lines.  The dotted curve gives the value of 
Im~T - T$^2$ - T$^{2}_{sf}$, where T$^{2}_{sf}$ is the spin-flip 
amplitude for C500.   
The real (imaginary) parts of single-energy solutions are plotted 
as filled (open) circles.  All amplitudes have been multiplied by a factor of 
10$^{3}$ and are dimensionless.  Plotted are the dominant partial-wave 
amplitudes:
(a) $^{3}P_{0}$ ($0^{+}$),
(b) $^{3}S_{1}$ ($1^{-}$),
(c) $^{3}P_{2}$ ($2^{+}$),
(d) $^{3}D_{2}$ ($2^{-}$),
(e) $^{3}D_{3}$ ($3^{-}$).}
\item
{Comparison of the $^3P_1S$ partial waves for $\pi d\to pp$ obtained in
the separate and combined fits. The real (imaginary) part of solution 
C500 is plotted as a solid (dashed) line. The purely real partial wave
from our separate analysis (SP96)\protect\cite{pidpp} 
is plotted as a dot-dashed line.}
\item
{Moduli of the partial-wave amplitudes for $\pi d\to pp$ from T$_{\pi}$ = 0 
to 500~MeV.  The solid and dashed curves give the 
amplitudes corresponding to the C500 and SP96\protect\cite{pidpp} solutions 
respectively.  Moduli of the single-energy solutions are plotted as 
filled circles.  All amplitudes have been multiplied by a factor of 10$^{3}$ 
and are dimensionless.  Only dominant partial-waves have been plotted:
(a) $^{1}S_{0} P$ ($0^{+}$),
(b) $^{3}P_{1} S$ ($1^{-}$),
(c) $^{1}D_{2} P$ ($2^{+}$),
(d) $^{3}P_{2} D$ ($2^{-}$),
(e) $^{3}F_{3} D$ ($3^{-}$).}
\end{list}

\eject

\begin{table}
\caption{Partial wave decomposition of $pp$, $\pi d$, and $N \Delta$ systems.}
\label{tbl1}
{\large
\vspace{0.3in}
\begin{center}

\begin{tabular}{cccc}
\tableline
$J^{P}$& $\pi d$          & $pp$                   & $N \Delta$              \\
\tableline
$0^{+}$&$^{3}P_{0}$       &$^{1}S_{0}$             &$^{5}D_{0}$              \\
\tableline
$0^{-}$&                  &$^{3}P_{0}$             &$^{3}P_{0}$              \\
\tableline
$1^{+}$&$^{3}P_{1}$       &                        &$^{3}S_{1}$, $^{3}D_{1}$ \\
       &$^{3}P_{1}$       &                        &$^{5}D_{1}$              \\
\tableline
$1^{-}$&$^{3}S_{1}$, $^{3}D_{1}$&$^{3}P_{1}$       &$^{3}P_{1}$              \\
       &$^{3}S_{1}$, $^{3}D_{1}$&$^{3}P_{1}$       &$^{5}P_{1}$, $^{5}F_{1}$ \\
\tableline
       &$^{3}P_{2}$, $^{3}F_{2}$&$^{1}D_{2}$       &$^{3}D_{2}$              \\
$2^{+}$&$^{3}P_{2}$, $^{3}F_{2}$&$^{1}D_{2}$       &$^{5}S_{2}$, $^{5}D_{2}$ \\
       &$^{3}P_{2}$, $^{3}F_{2}$&$^{1}D_{2}$       &$^{5}D_{2}$, $^{5}G_{2}$ \\
\tableline
$2^{-}$&$^{3}D_{2}$       &$^{3}P_{2}$, $^{3}F_{2}$&$^{3}P_{2}$, $^{3}F_{2}$ \\
       &$^{3}D_{2}$       &$^{3}P_{2}$, $^{3}F_{2}$&$^{5}P_{2}$, $^{5}F_{2}$ \\
\tableline
$3^{+}$&$^{3}F_{3}$       &                        &$^{3}D_{3}$, $^{3}G_{3}$ \\
       &$^{3}F_{3}$       &                        &$^{5}D_{3}$, $^{5}G_{3}$ \\
\tableline
$3^{-}$&$^{3}D_{3}$, $^{3}G_{3}$&$^{3}F_{3}$       &$^{3}P_{3}$, $^{3}F_{3}$ \\
       &$^{3}D_{3}$, $^{3}G_{3}$&$^{3}F_{3}$       &$^{5}P_{3}$, $^{5}F_{3}$ \\
       &$^{3}D_{3}$, $^{3}G_{3}$&$^{3}F_{3}$       &$^{5}F_{3}$, $^{5}H_{3}$ \\
\tableline
       &$^{3}F_{4}$, $^{3}H_{4}$&$^{1}G_{4}$       &$^{3}G_{4}$              \\
$4^{+}$&$^{3}F_{4}$, $^{3}H_{4}$&$^{1}G_{4}$       &$^{5}D_{4}$, $^{5}G_{4}$ \\
       &$^{3}F_{4}$, $^{3}H_{4}$&$^{1}G_{4}$       &$^{5}G_{4}$, $^{5}I_{4}$ \\
\tableline
$4^{-}$&$^{3}G_{4}$       &$^{3}F_{4}$, $^{3}H_{4}$&$^{3}F_{4}$, $^{3}H_{4}$ \\
       &$^{3}G_{4}$       &$^{3}F_{4}$, $^{3}H_{4}$&$^{5}F_{4}$, $^{5}H_{4}$ \\
\tableline
\end{tabular}
\end{center}
}
\end{table}

\eject

\begin{table}
\caption{Comparison of the combined analysis (C500) and our previous (separate) 
analyses.  WI96 for $pp\to pp$ \protect\cite{pp}, SM94 for 
$\pi d\to \pi d$ \protect\cite{pid}, 
and SP96 for $\pi d\to pp$ \protect\cite{pidpp}.  The relevant energy ranges 
are: T$_{\pi}$ = 0--500~MeV, T$_{p}$ = 288--1290~MeV, and $\protect\sqrt{s}$ = 
2015--2440~MeV.}
\label{tbl2}
{\large
\begin{center}
\begin{tabular}{ccc}
\tableline
Reaction          & Separate      & Combined      \\
                  & $\chi^2$/Data & $\chi^2$/Data \\
\tableline
$pp\to pp$        & $17380/10496$ & $17380/10496$ \\
$\pi d\to \pi d$  &  $2745/1362$  &  $2418/1362$  \\
$\pi d\to pp$     &  $7716/4787$  &  $7570/4787$  \\
\tableline
\end{tabular}
\end{center}
}
\end{table}

\eject

\begin{table}
\caption{Comparison of single-energy (binned) and energy-dependent combined
analyses of $\pi d$ elastic scattering data.  $N_{prm}$ is the number 
of parameters varied in the single-energy fits. $\chi^2_E$ is due to the
energy-dependent fit (C500) taken over the same energy interval.}
\label{tbl3}
{\large
\begin{center}
\begin{tabular}{ccccccc}
T$_{\pi}$~(MeV)&Range~(MeV)&$N_{prm}$&$\chi^2$/data&$\chi^2_E$&&\\
\tableline
  65 & $  58.0 -  72.0 $ &  2 & 106/54  &  102 &&\\
  87 & $  72.0 -  92.0 $ &  6 &  20/24  &   21 &&\\
 111 & $ 107.5 - 125.2 $ & 10 &  68/82  &   66 &&\\
 125 & $ 115.0 - 134.0 $ & 12 & 155/170 &  184 &&\\
 134 & $ 124.0 - 142.8 $ & 14 & 315/258 &  344 &&\\
 142 & $ 133.0 - 152.0 $ & 16 & 356/284 &  397 &&\\
 151 & $ 141.0 - 160.6 $ & 16 & 193/154 &  216 &&\\
 182 & $ 174.0 - 189.5 $ & 18 & 302/168 &  396 &&\\
 216 & $ 206.0 - 220.0 $ & 18 & 158/99  &  200 &&\\
 230 & $ 220.0 - 238.0 $ & 18 &  64/53  &  111 &&\\
 256 & $ 254.0 - 260.0 $ & 16 & 132/125 &  185 &&\\
 275 & $ 270.5 - 284.4 $ & 16 &  22/40  &   42 &&\\
 294 & $ 284.4 - 300.0 $ & 16 & 267/132 &  324 &&\\
\end{tabular}
\end{center}
}
\end{table}

\eject

\begin{table}
\caption{Comparison of single-energy (binned) and energy-dependent combined
analyses of $\pi d\to pp$ reaction data.  $N_{prm}$ is the number of 
parameters varied in the single-energy fits. $\chi^2_E$ is due to the
energy-dependent fit (C500) taken over the same energy interval.}
\label{tbl4}
{\large
\begin{center}
\begin{tabular}{ccccccc}
T$_{\pi}$~(MeV)&Range~(MeV)&$N_{prm}$&$\chi^2$/data&$\chi^2_E$&&\\
\tableline
  25 & $  12.8 -  37.4 $ & 10 & 527/241 &  542 &&\\
  50 & $  37.6 -  60.7 $ & 12 & 188/168 &  205 &&\\
  75 & $  62.9 -  87.3 $ & 14 & 590/426 &  628 &&\\
 100 & $  91.0 - 114.0 $ & 14 &1263/611 & 1379 &&\\
 125 & $ 113.8 - 137.1 $ & 16 & 729/512 &  756 &&\\
 150 & $ 140.0 - 162.0 $ & 20 & 743/630 &  792 &&\\
 175 & $ 165.0 - 187.3 $ & 22 & 343/280 &  426 &&\\
 200 & $ 191.3 - 210.3 $ & 20 & 120/193 &  153 &&\\
 225 & $ 217.9 - 235.9 $ & 22 & 217/229 &  291 &&\\
 250 & $ 238.9 - 262.0 $ & 22 & 595/483 &  685 &&\\
 275 & $ 264.9 - 285.1 $ & 22 & 204/109 &  280 &&\\
 300 & $ 291.6 - 307.4 $ & 24 & 198/212 &  235 &&\\
 325 & $ 318.9 - 330.0 $ & 24 & 142/161 &  234 &&\\
 350 & $ 341.4 - 360.3 $ & 24 & 201/185 &  233 &&\\
 375 & $ 371.4 - 375.7 $ & 24 &  32/26  &   42 &&\\
 400 & $ 390.0 - 400.0 $ & 24 &  19/28  &   34 &&\\
 425 & $ 417.0 - 420.0 $ & 24 &  50/28  &   55 &&\\
 450 & $ 437.6 - 456.5 $ & 22 & 122/48  &  231 &&\\
 475 & $ 473.8 - 487.4 $ & 22 &  24/24  &   39 &&\\
 500 & $ 495.9 - 506.5 $ & 22 &  49/45  &  281 &&\\
\end{tabular}
\end{center}
}
\end{table}

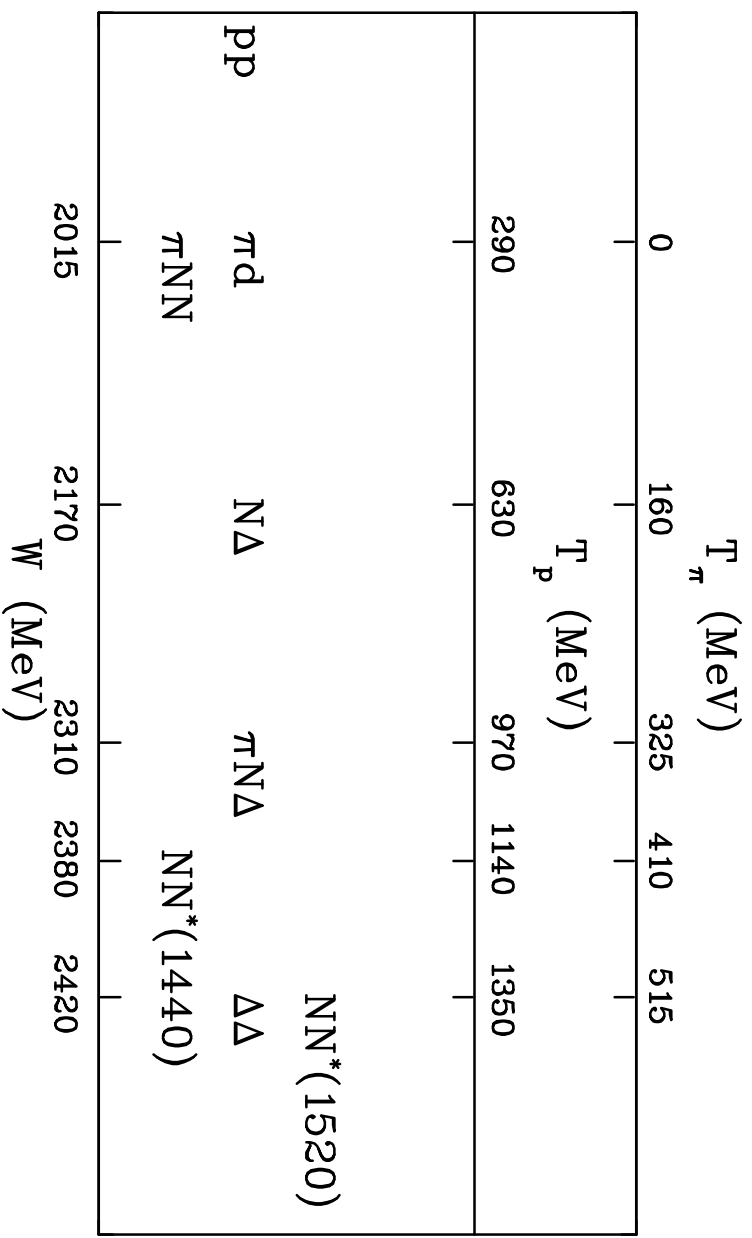
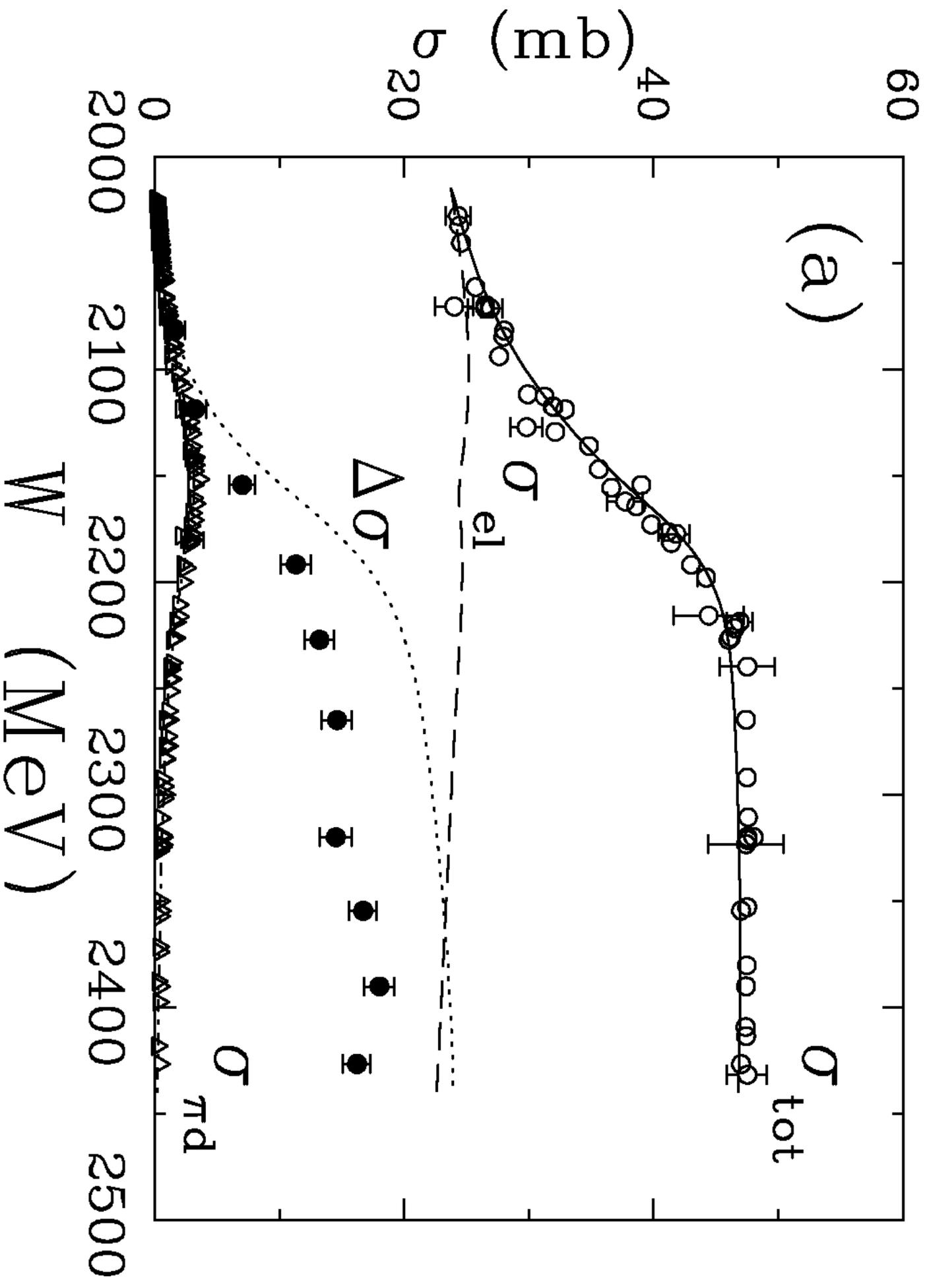
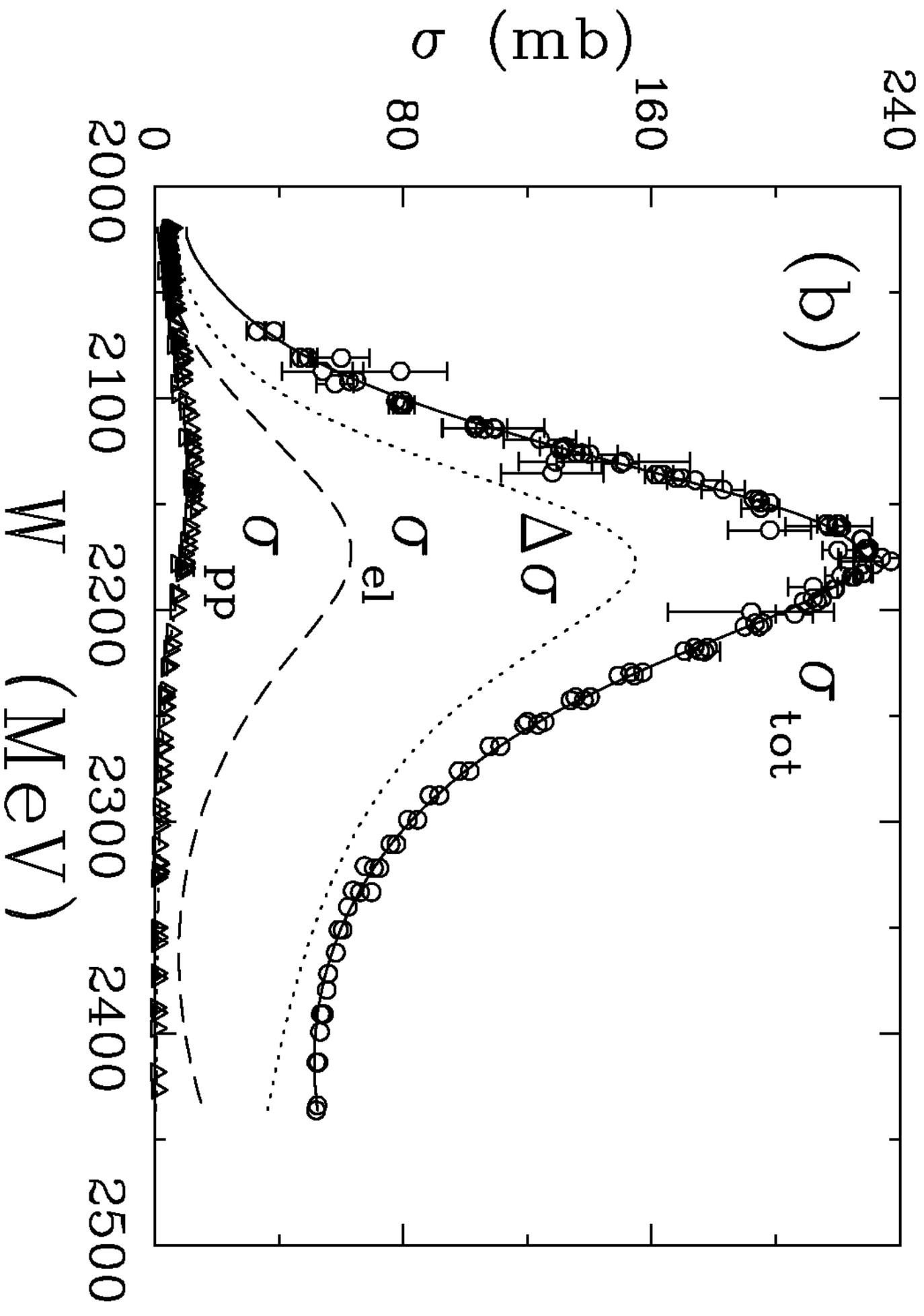
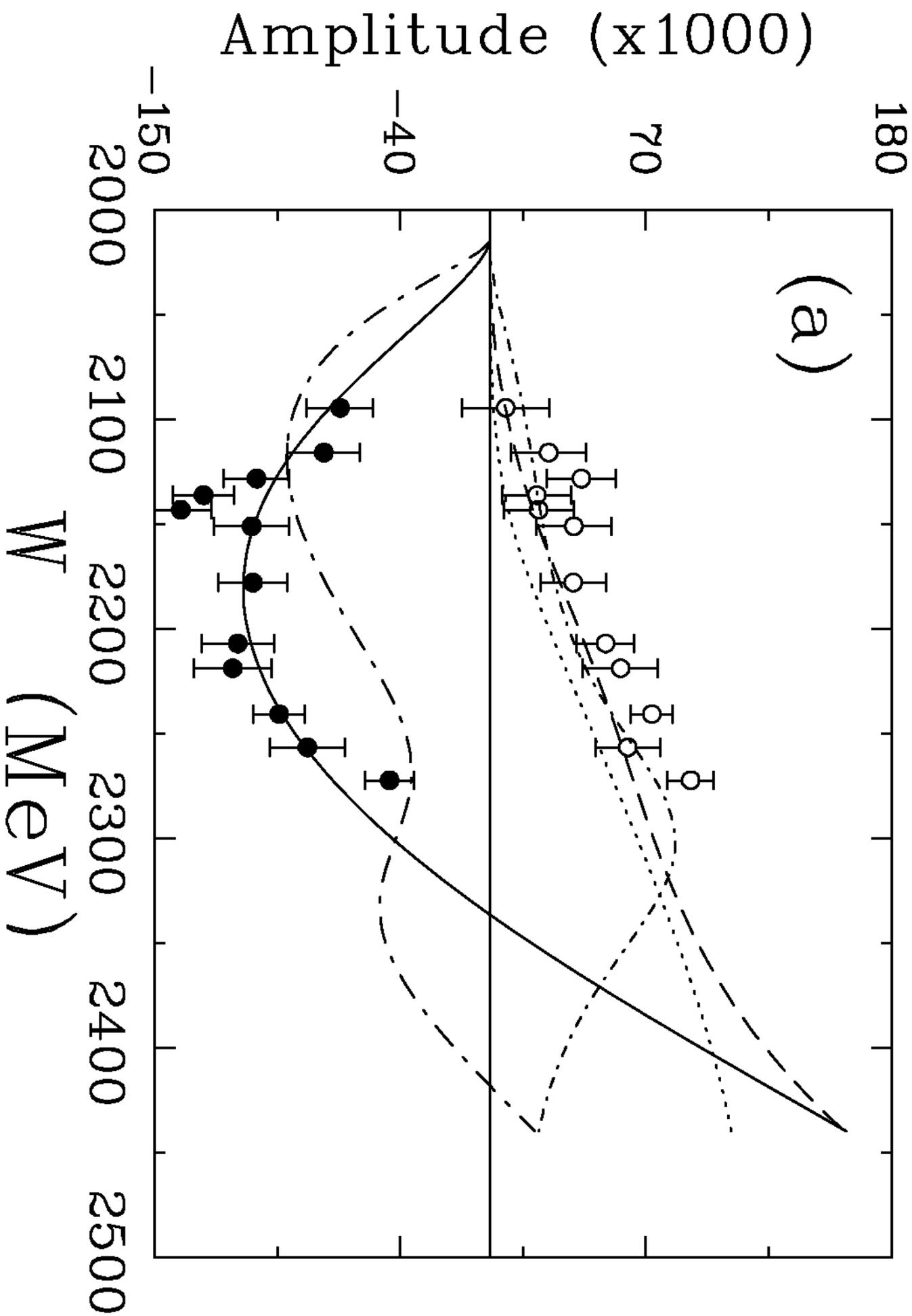
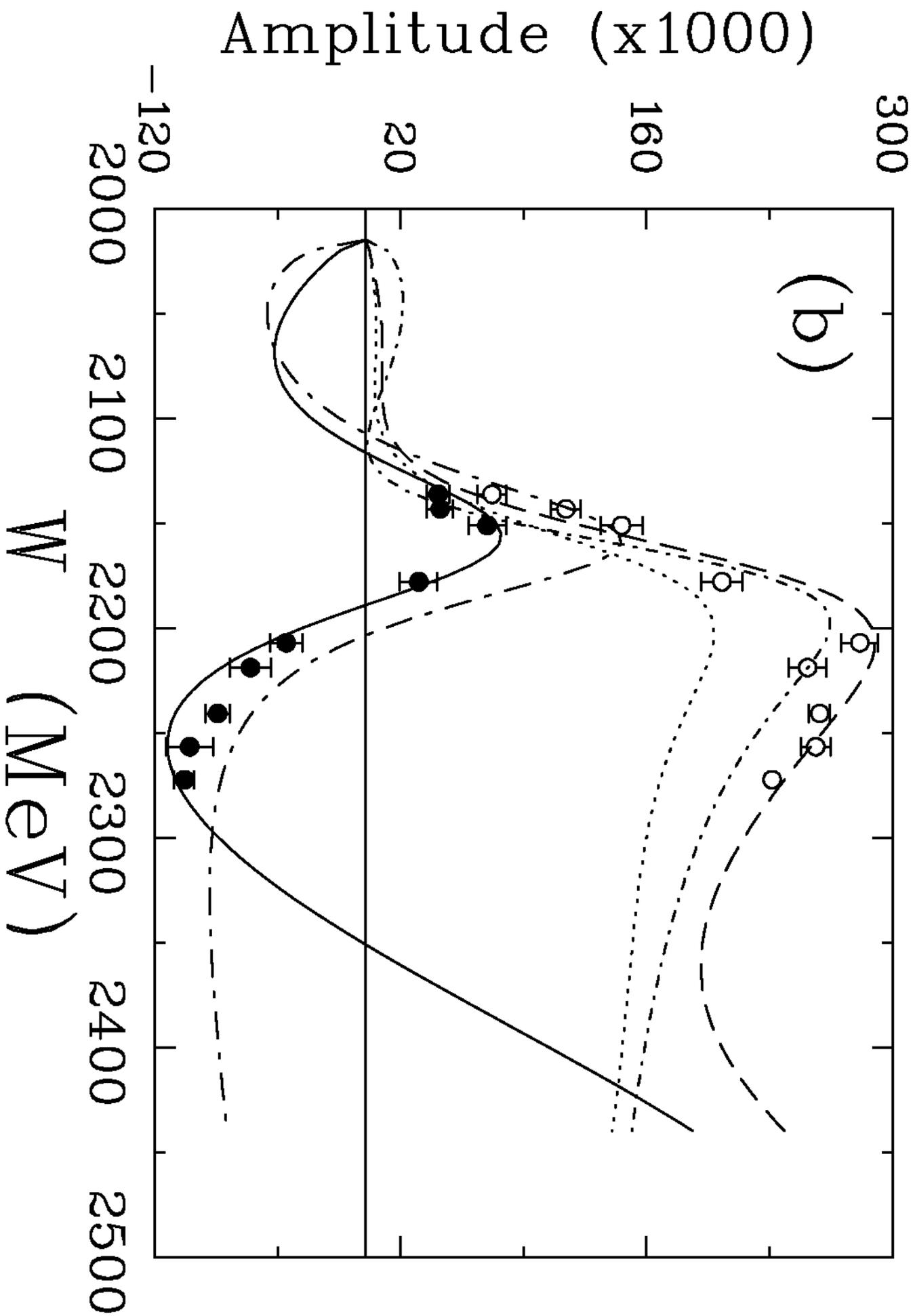
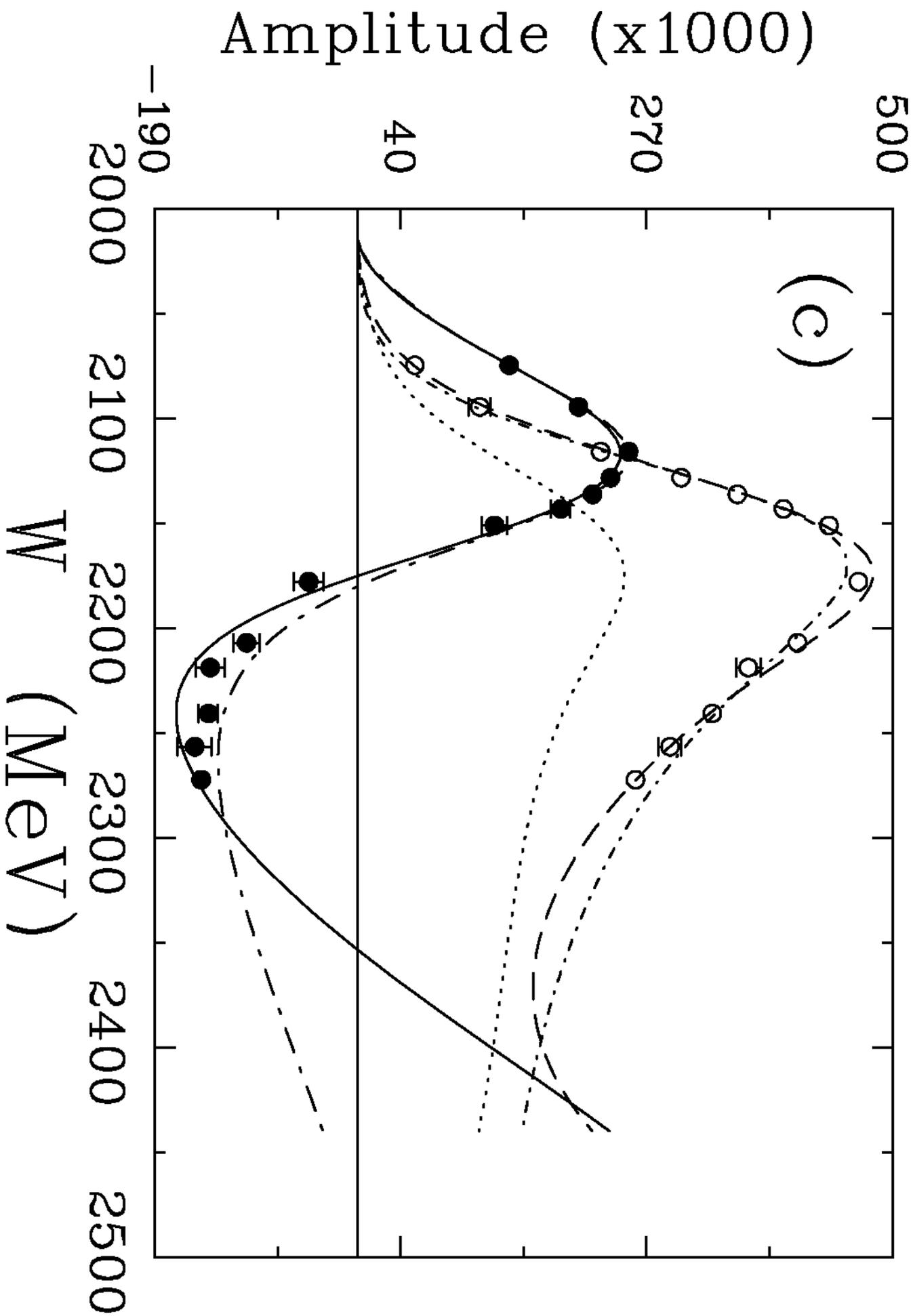
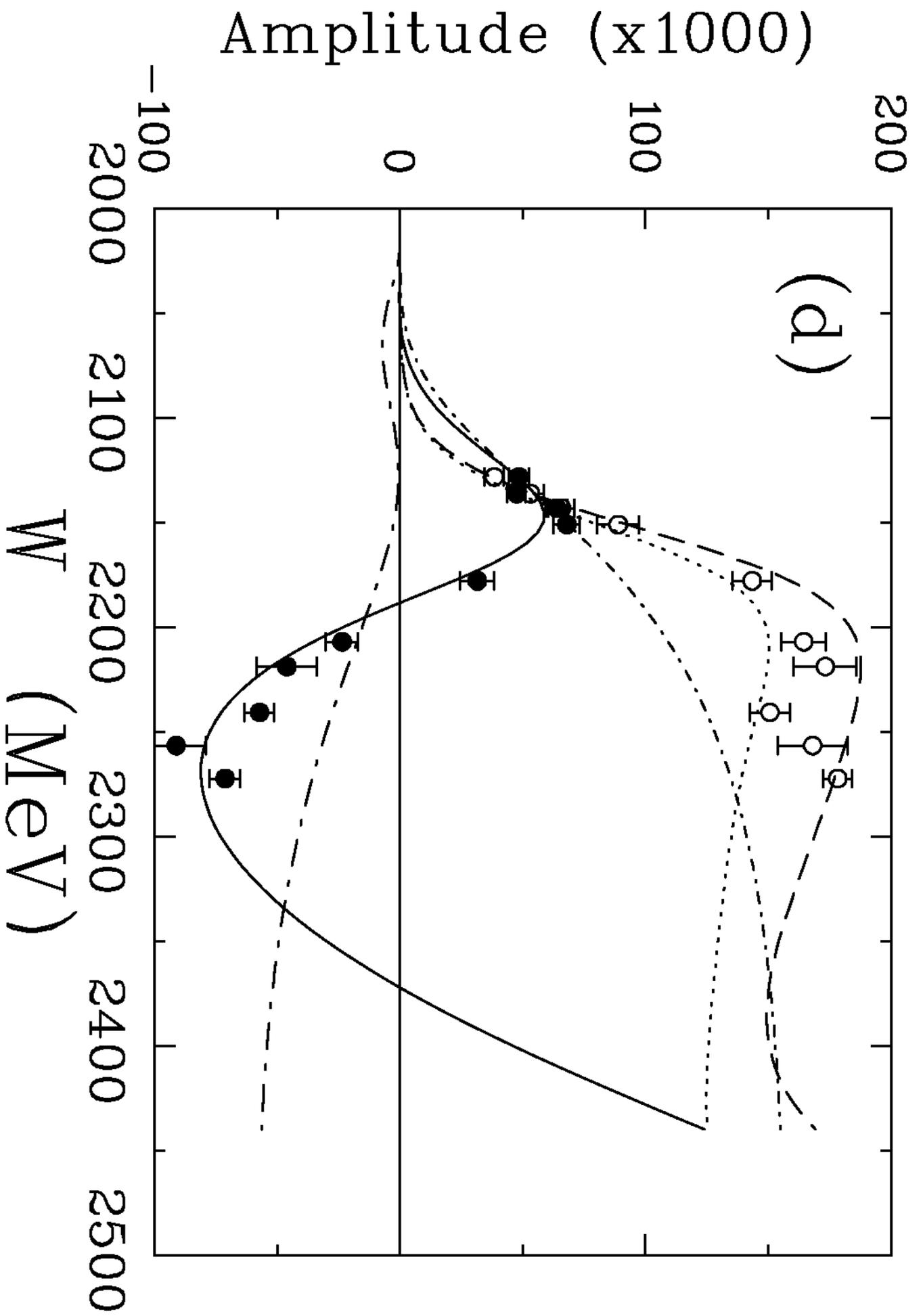
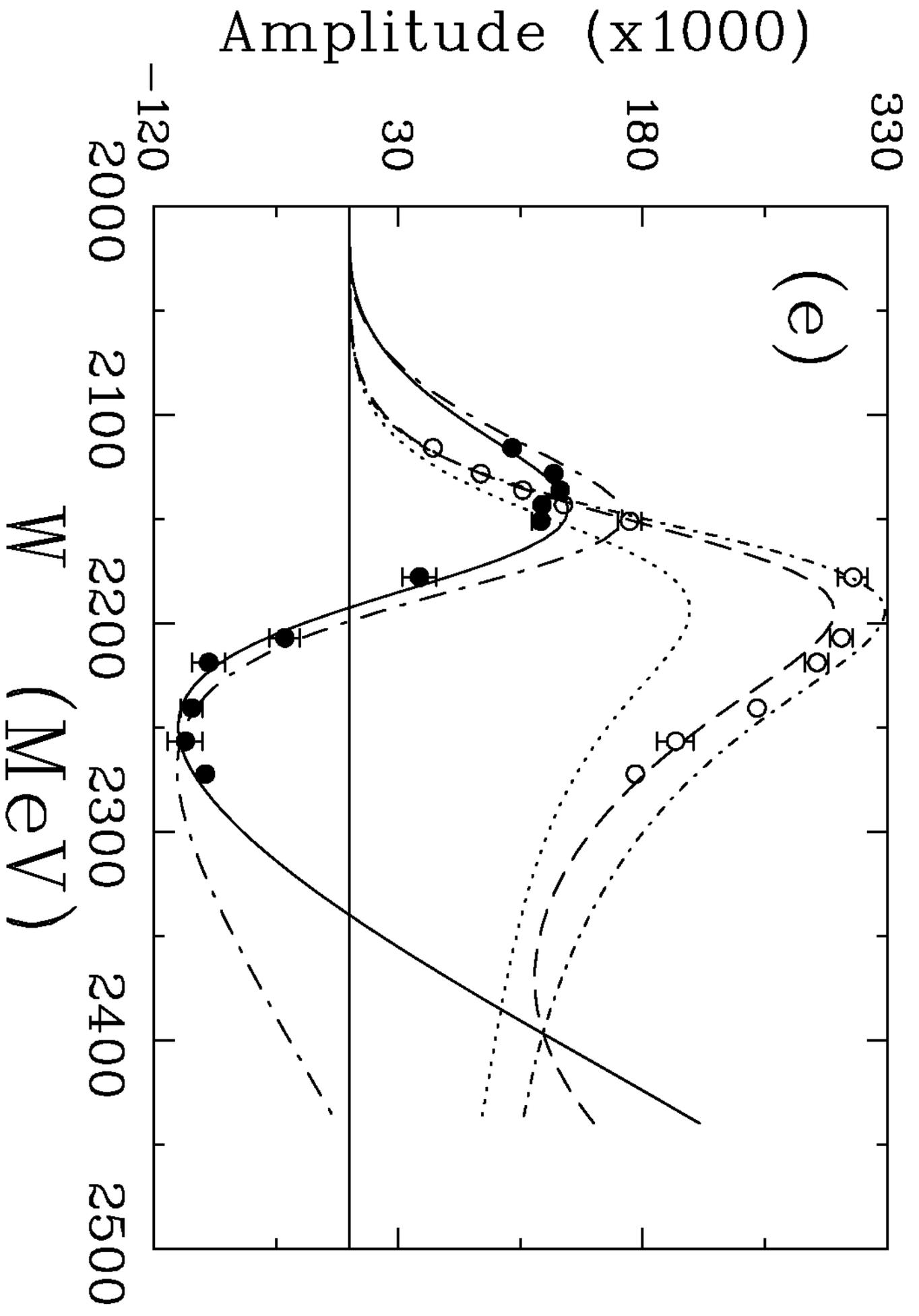
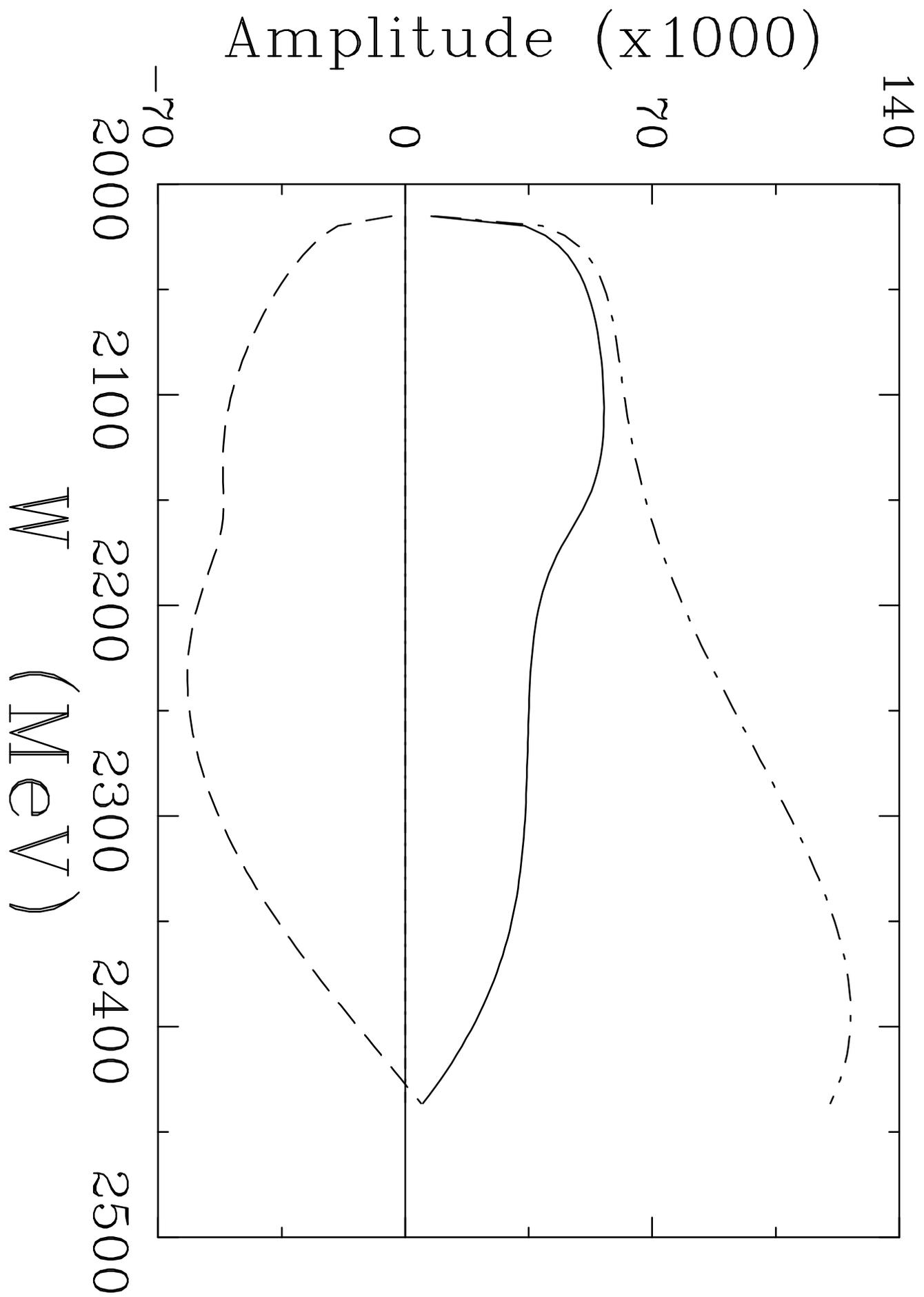
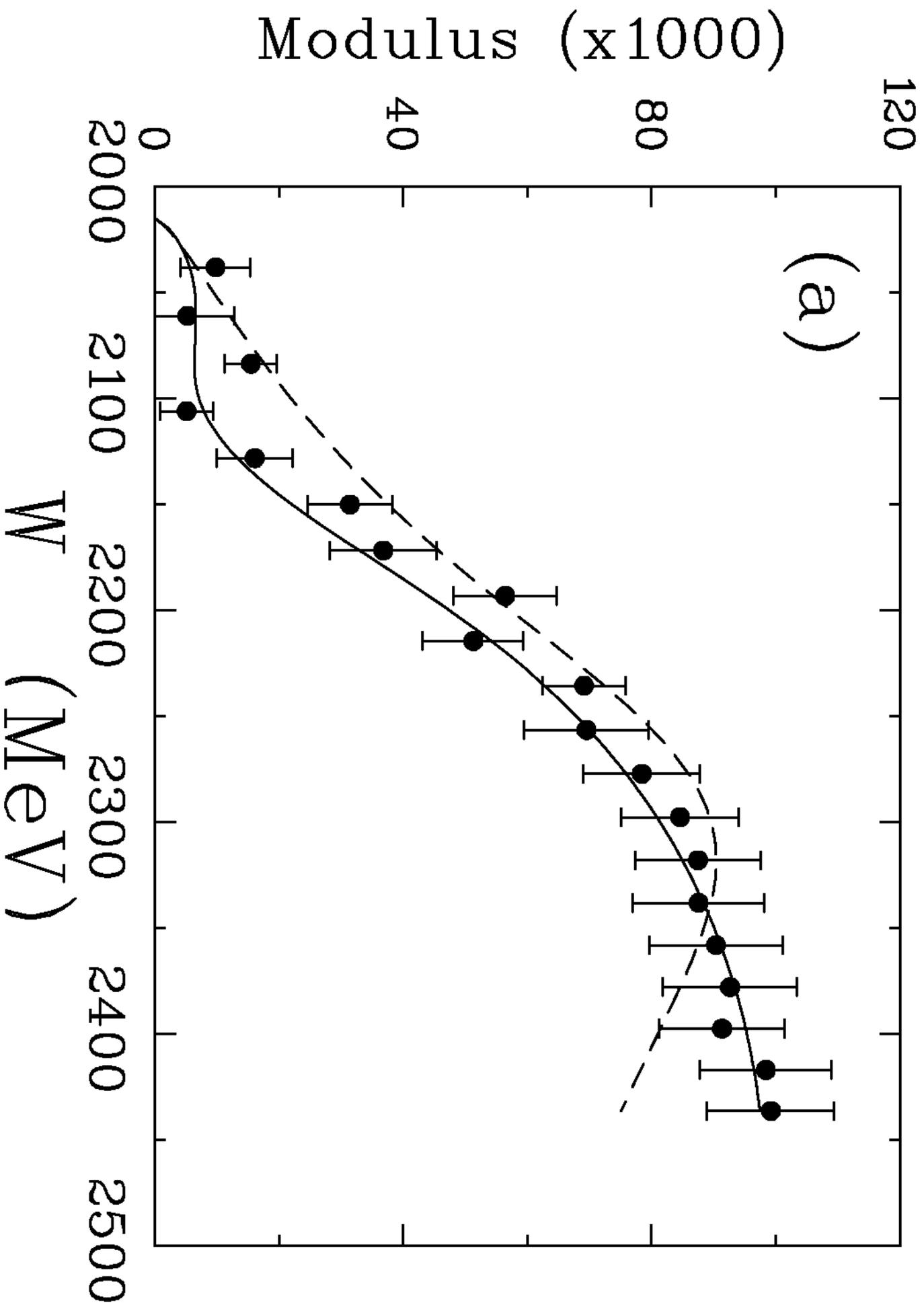
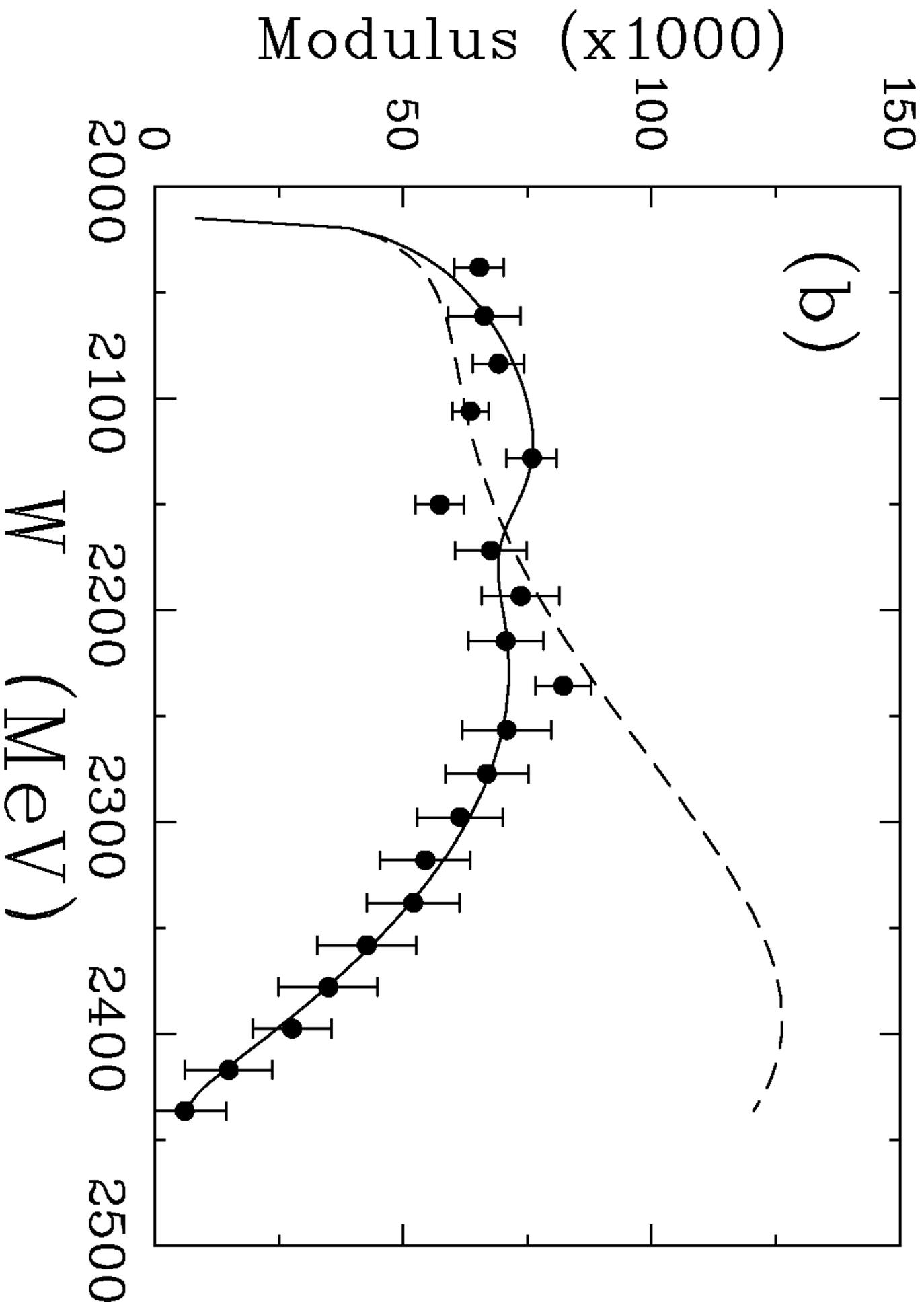
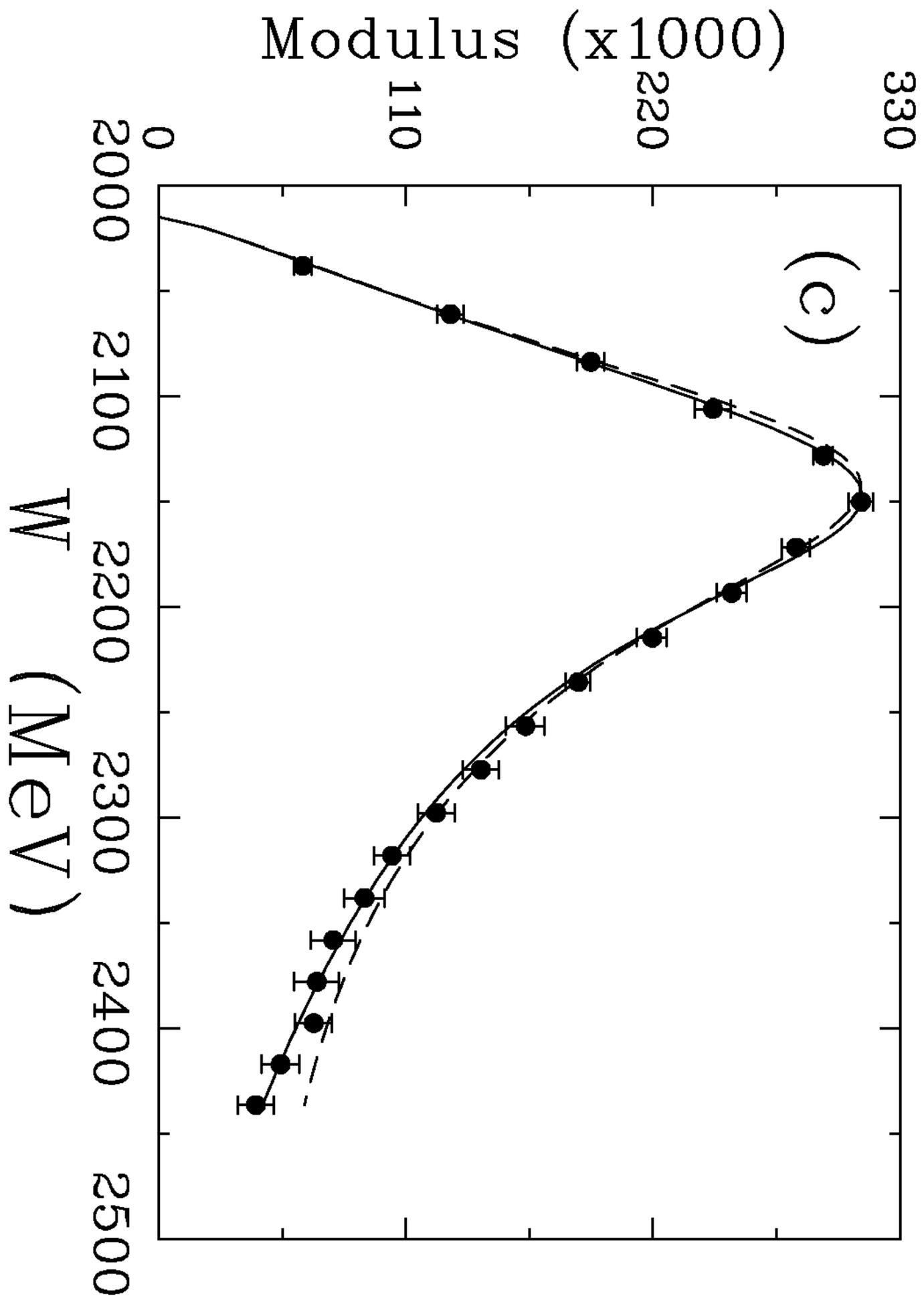
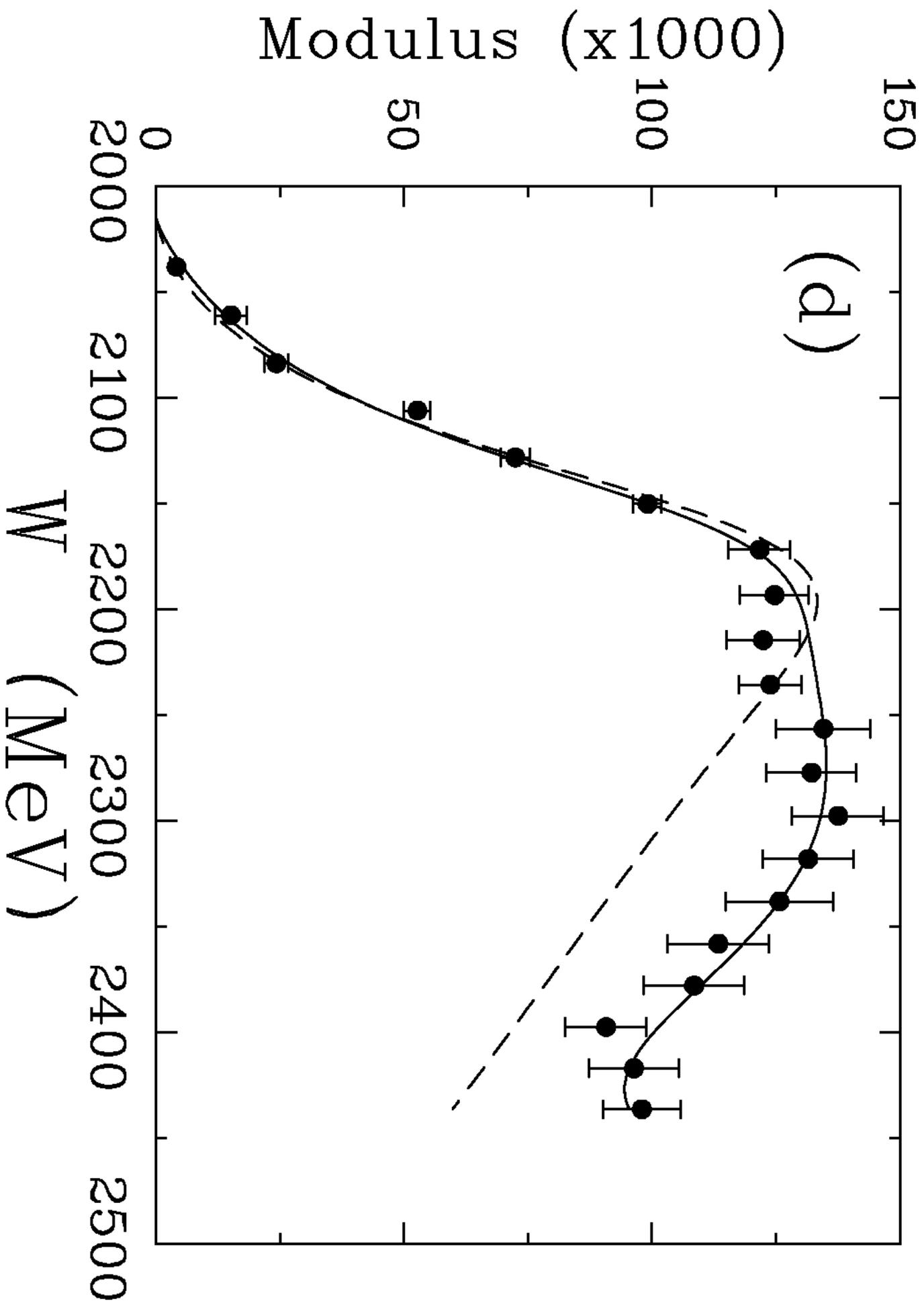
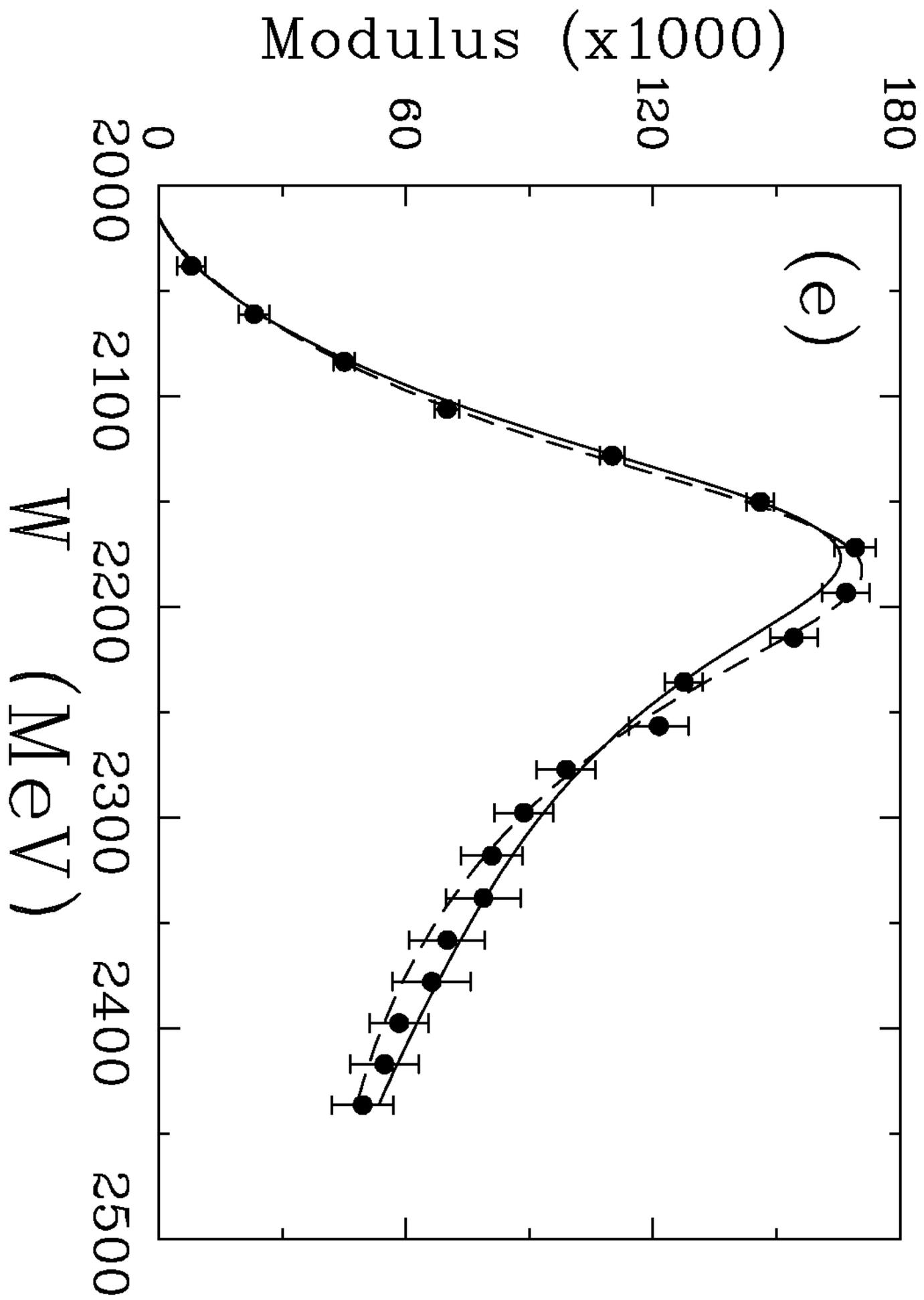
\end{document}